# Toward Generating Synthetic CT Volumes using a 3D-Conditional Generative Adversarial Network

Jayalakshmi Mangalagiri[1*], David Chapman[1], Aryya Gangopadhyay[1], Yaacov Yesha[1], Joshua Galita[1], Sumeet Menon[1], Yelena Yesha[1], Babak Saboury[1,2], Michael Morris[1,2,3], Phuong Nguyen[1,4]

[1]*University of Maryland, Baltimore County*, Baltimore, MD, USA, [2]*National Institutes of Health Clinical Center*, Bethesda, MD, USA, [3]*Networking Health*, Glen Burnie MD, USA, [4]*OpenKneck Inc*, Halethorpe, MD, USA

*corresponding author: jmangal1@umbc.edu*

*Abstract*—We present a novel conditional Generative Adversarial Network (cGAN) architecture that is capable of generating 3D Computed Tomography scans in voxels from noisy and/or pixelated approximations and with the potential to generate full synthetic 3D scan volumes. We believe conditional cGAN to be a tractable approach to generate 3D CT volumes, even though the problem of generating full resolution deep fakes is presently impractical due to GPU memory limitations. We present results for autoencoder, denoising, and depixelating tasks which are trained and tested on two novel COVID19 CT datasets. Our evaluation metrics, Peak Signal to Noise ratio (PSNR) range from 12.53 - 46.46 dB, and the Structural Similarity index (SSIM) range from 0.89 to 1.

Type of the submission: "short paper"

This paper is being submitted to CSCI-ISHI for : Health Informatics and Medical Systems.

## I. Introduction

Generative Adversarial Networks (GANs) are popular for many medical imaging informatics tasks involving CT scans. Yet to the best of our knowledge no GAN algorithms are capable of generating full-sized full-resolution synthetic CT scans from white noise. Rather, GANs for CT operate on either individual axial slices or on small 3D regions of interest (RoI). This is because most CNN generators require roughly ~10 GB of GPU memory in order to generate a single 512x512 axial slice. Naively scaling a GAN to generate a coherent full resolution CT scan with 100-200 axial slices would take two orders of magnitude more memory than is available in present day GPU hardware, making this problem impractical for the foreseeable future.

There are many potential use cases for generating full-resolution CT scan images, but one notable use case is to fabricate *deep-fakes* i.e. highly realistic images of non-existent patients presenting a condition. There are many large clinical datasets that cannot be easily shared due to privacy concerns. If *deep-fakes* were possible, then realistic images could be fabricated of patients that do not exist, thereby eliminating privacy concerns and providing more needed data.

Coronavirus is a widespread virus that has caused a global pandemic [2,3,8,11], and Deep learning algorithms have shown an ability to detect pneumonia symptoms even in patients with early stages of disease [7,13]. Yet the performance of these algorithms is severely limited by training data volumes of publicly available datasets [1, 9, 14 - 19]. Furthermore, recent works have shown that COVID19 images can benefit from the introduction of synthetic GAN imagery, even if these images are not of diagnostic quality [1, 19]. We have IRB approval to work with 1049 CT scans for the purposes of developing *synthetic data* that can be made publicly available. If successful, our dataset would be the largest publicly available COVID19 imaging dataset even though none of the patients in the public version would be actual people.

We present a conditional GAN (cGAN) approach to improve the resolution of 3D volume blocks and demonstrate the ability to autoencode full-resolution COVID19 CT scans, as well as contiguously depixelate / denoise the lung region of the image. At present the cGAN architecture can synthesize full resolution imagery from an approximate image condition. In future work we intend to employ this architecture iteratively in order to fabricate full resolution CT scans from white noise.

## II. Background and related work

A GAN network consists of a Generator G and a Discriminator D. Generator G(Z) generates images from random noise Z, whereas Discriminator D(X), estimates the probability that an image comes from the real dataset as opposed to generated[2]. The minimax loss equation (1) is as follows [8],

$$min_G min_D V(D,G) = E_X[ln\, D(X)] + E_Z[ln\,(1 - D(G(Z)))] \quad (1)$$

cGAN is a conditional GAN model in when both the Generator as well as Discriminator are conditioned on auxiliary information Y. Y is therefore an additional input layer, and the conditional minimax loss is as follows (2) [8],

$$min_G min_D V(D,G) = E_X[ln\, D(X|Y)] + E_Z[ln\,(1 - D(G(Z|Y)))] \quad (2)$$

cGAN models are often image-to-image translation problems [1, 4]. GAN models have been heavily used in medical imaging applications [19]. But patient CT scans can have resolution from 512x512x600 to 1024x1024x600 or larger leading to hundreds of millions of voxel radiodensities in Hounsfield units (HU). CNN architectures for generating a single 512x512 slice typically require ~10GB of GPU memory, and thus generating an entire 3D CT scan volume by GAN is presently impractical, even though it is possible to generate either low-resolution CT scans, or high resolution (Regions of Interest). Mirsky et al have shown that it is possible for an attacker using deep learning to generate realistic nodules for injecting into 3D CT-scans or removing nodules from 3D CT scans using a cGAN [1]. We have used a similar approach of using 3D high resolution sub voxels as training samples for condition GAN models. We divide full 3D CT scans into non overlapping high resolution sub voxels and use all of them for training. Khalifa et al. presented a fine-tuned deep transfer learning for a limited dataset by a pneumonia chest x-ray detection based on generative adversarial networks (GAN) at low resolution [2]. Menon et al. have shown that GANs with mean teacher can generate realistic fake X-rays to improve the accuracy of COVID-19 screening classification, although X-rays have far lower memory requirements than CT-scans [6]. Shan et al. developed a deep learning based segmentation system for assessing the COVID-19 quantitatively [12].

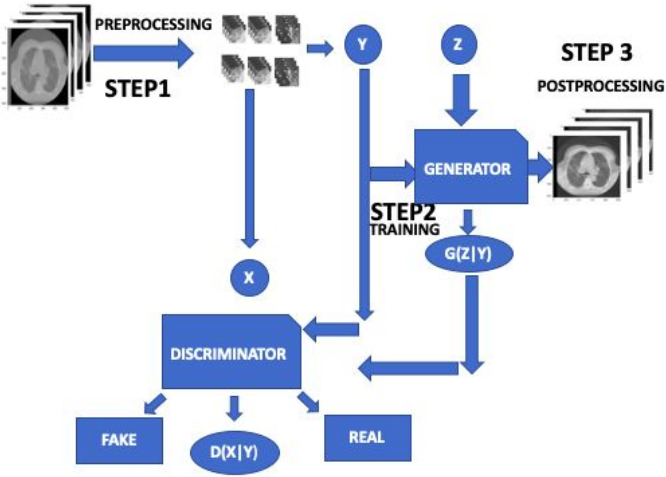

Fig. 1. Training workflow cGAN framework including architectural diagram

### III. METHODOLOGY

#### A. Conditional GAN models

We employ a conditional GAN (cGAN) model to generate synthetic data from an approximation of 3D full CT scans as subdivided into 32x32x32 voxel blocks [4][5]. The description of cGAN is shown in equation (2) where input X are the non overlapping subdivided 32x32x32 blocks, and condition Y can be either the original, noisy, or pixelated voxels. For the autoencoder, the condition is identical to the original voxels. The condition for the noisy voxels is created by introducing Poisson noise to the original voxels. The pixelated voxels are created by downsampling voxels in X by half of its resolution with a nearest neighbor filter. The pixelated condition can be considered as a simulation of low resolution or blur input. A random noise vector Z is also drawn from a uniform distribution feeding to the generator initially. The Discriminator is supplied with a condition Y, fake $G(Z|Y)$ as well as target X. The proposed cGAN architecture employs a 3D U-net as the Generator architecture and also includes a pixel wise L1 loss. A training workflow diagram showing the cGAN architecture is seen in Fig 1. Each of the steps in the Fig.1. is briefly explained in the following sections.

#### B. Description of input datasets

We make use of 1049 de-identified CT scan Covid-19 infected cases from two novel data sources. Dataset1 has CT 49 exams of COVID19 infected patients each with multiple series. We use series 3 which presents 70 to 90 slices per scan with slice thickness ranging from 3.75mm to 5mm at 512x512 resolution. Dataset2 has 1000 exams of COVID-19 infected patients each also with 6 different series. The number of slices ranges approximately from 20 to 300. The slice thickness varies from 3.5mm to 7.5mm and resolution is 512x512 pixels. Due to limited computational time, we make use of 37 patients CT scans from these 2 Datasets, splitting 80% for training (31 patients) and 20% for testing (6 patients). These 37 exams consist of ~3000 512x512 axial slices in total.

#### C. Preprocessing and Training

Each CT-scan image is preprocessed by a) dividing the image into 32x32x32 voxel , b) equalization / normalization, and c) augmentation. Dividing into voxels of 32x32x32 resolution is necessary for the Generator's feature vectors to reasonably fit into GPU memory. Histogram equalization and intensity normalization convert radiodensities to a [-1, -1] range increasing contrast of textural features. Histogram and normalization parameters are recorded to enable results to be converted back to original scale in postprocess. We employ an augmentation factor of 34, by incorporating translation, mirroring, and rotations. This yields a total training volume of 170,748 voxels (162 voxels/ scan × 31 scans × 34 augmentations) and similarly a testing volume of 33,048 voxels.

#### D. Training cGAN model

The cGAN model was trained for 100 epochs with a batch size of 50 samples. We evaluate this model using two experiments to ensure the ability of cGAN to replicate realistic imagery. The first experiment is a deep autoencoder in which the cGAN was tasked to generate synthetic images that replicate the original images voxel by voxel. Training of cGAN for this experiment required 8 days and 7 hours to complete using a 32-Core AMD with 32GB RAM and 2 Geforce 2080 Ti 11GB GPUs. Our validation accuracy achieved 99-100% upon completion of training. In our second experiment we performed pulmonary depixelization, denoising, and autoencoding. To reconstruct full scans we perform reverse preprocessing and then image-stitching to combine the predicted voxels to a full CT scan image with resolution of 512x512xZ where Z is the number of axial slices.

IV. EXPERIMENTAL RESULTS

We perform two sets of experiments, the first is a full size CT autoencoder task, and the second is a comparison of the autoencoder versus depixelator versus denoising task over the pulmonary regions. The Autoencoder achieved accurate results with slices presented in Fig 2. The cGAN generated (Fig 2.c) CT scans and original contrast enhanced (Fig 2.b) CT scans show very little difference as seen in the subtraction image (Fig 2.d). There is some noise inside the lung showing minor differences. Most of the noticeable differences are in the soft tissue and bone areas outside of the lung. Most importantly there are no obvious artifacts as seen in the cGAN generated images.

Figures 3, 4, and 5 show the resulting images of a 32x32 region for the autoencoder, denoising, and depixelating tasks respectively. We see in Fig 3 that the autoencoder produces very realistic images, although there is reduced contrast over a bone feature in (Fig 3 mid) and minor blurring of textural details in the lung region (Fig 3 right). Fig 4 shows that the denoising task is able to improve the image quality and reduce noise although some blurring of contrast is also apparent. The depixelating results (Fig 5) show that the generated images improve textural detail over the pixelated condition.

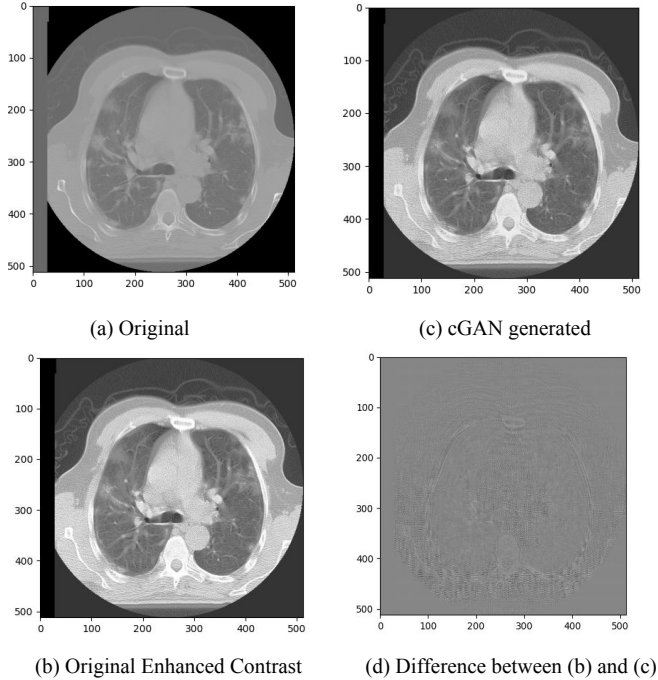

(a) Original

(c) cGAN generated

(b) Original Enhanced Contrast

(d) Difference between (b) and (c)

Fig. 2. Original vs cGAN generated and its differences.

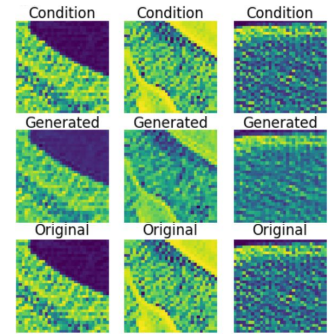

Fig 3. Close up comparison of condition (top) generated (mid) and original (bottom) 32x32 regions for Autoencoder task.

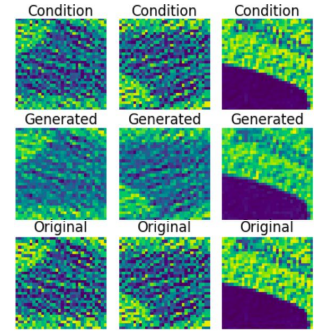

Fig 4. Close up comparison of condition (top) generated (mid) and original (bottom) 32x32 regions for Denoising task.

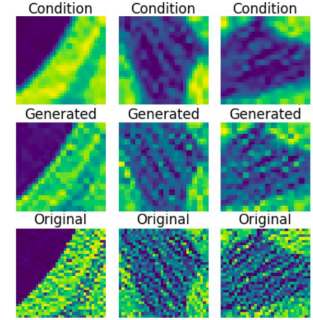

Fig 5. Close up comparison of condition (top) generated (mid) and original (bottom) 32x32 regions for Depixelating task.

### A. Quantitative Evaluation Metrics

We employ the Peak signal-to-noise ratio (PSNR) and Structural Similarity Index (SSIM) as evaluation metrics [10] with equations as follows,

$$PSNR = 10 \, log_{10}(\frac{I}{MSE})  \quad (3)$$

$$SSIM = \frac{(2\mu_x\mu_y + c_1)(2\sigma_{xy} + c_2)}{(\mu_x^2 + \mu_y^2 + c_1)(\sigma_x^2 + \sigma_y^2 + c_2)} \quad (4)$$

TABLE I PSNR AND SSIM FOR AUTOENCODER TASK

| PATIENT | PSNR (units = dB) | SSIM |
|---|---|---|
| 1 | 31.02 | 0.91 |
| 2 | 31.92 | 0.97 |
| 3 | 15.08 | 0.95 |
| 4 | 24.84 | 0.97 |
| 5 | 31.53 | 0.98 |

| 6 | 12.53 | 0.89 |

Table I shows the PSNR and SSIM performance for the cGAN autoencoder task. SSIM is a metric that estimates the perceived quality of images and videos by measuring the similarity between the original and the recovered images [10]. The range of SSIM is -1 to +1 where +1 occurs only when the two images are identical [21]. The peak signal-to-noise ratio (PSNR) although imperfect is designed to approximate human perception of the image reconstruction quality [10]. The PSNR values vary from 30-50 db for a 8-bit data representation and 60-80 db for a 16-bit data representation [10].

In Table I, the PSNR and SSIM are calculated by averaging over all slices of the full CT image generated from cGAN vs original full CT images. PSNR ranged from 12.53 to 46.46 dB for patients 1-6, and SSIM ranged from 0.89 to 0.97. The autoencoder achieves above 0.95 SSIM over 4 of 6 patients, and achieves above 31 dB PSNR in 3 of the 6 patients observed. These results show that the cGAN achieves good performance in preserving image quality although there is some variation with patients 3 and 6 in a lower range of 12.53 - 15.08 dB. This variance can be explained because we observe that the autoencoder occasionally produces artifact blocks which can lead to increased variance in PSNR on a patient basis when dealing with differences in patients CT scans (slice thickness, contrast levels and acquired conditions). For example, we observed CT scans with inclined images.

TABLE II PSNR AND SSIM FOR AUTOENCODER DENOISING AND DEPIXELATING OVER PULMONARY REGION

| ConditionGAN | PSNR (Units= dB) | SSIM |
|---|---|---|
| cGAN Autoencoder | 46.46 | 1 |
| cGAN Denoised | 46.34 | 1 |
| poisson noise Image | 27.29 | 0.99 |
| cGAN Depixelated | 37.26 | 0.99 |
| pixelated Image | 39.5 | 0.99 |

Table II shows a comparison of 3 tasks (Autoencoder, Denoising, and Depixelating) using the same set of 31 patients for training 6 patients for testing but limited to 4 voxel (32x32x32 each) over pulmonary area in order to reduce the computation burden. Table II shows that our cGAN can predict a high PSNR for denoising of 46.34 dB which is comparable to the autoencoder task with PSNR of 46.46 dB. For comparison, the image with Poisson noise added has much lower PSNR of 27.29 dB than the cGAN Denoised of 46.34 dB. We observe however that the cGAN depixelated image achieves a slightly lower PSNR of 37.26 dB as compared with the pixelated image of 39.5 dB.

### B. Qualitative Evaluation Metrics

Three human volunteers were asked to identify 20 pairs of Real/Generated image slices from the Autoencoder task at the full size of 512x512 pixels. The generated images were identified 50% correctly which is indistinguishable from random guessing. Of the images identified correctly the most noticeable artifacts were due to very small size differences in structures outside of the lung region. Contrast differences were also observed even though explicit normalization was removed. Nevertheless, the cGAN generated images were visually observed to reproduce variations in slice thickness, and reconstruction kernels without apparent artifacts in the lung area.

### V. CONCLUSION

We present a novel conditional Generative Adversarial Network (cGAN) architecture that is capable of generating 3D Computed Tomography scan blocks from noisy and/or pixelated approximations and with the potential to generate full synthetic 3D scan volumes in future work. Generated image quality evaluation shows Peak Signal to Noise Ratio (PSNR) ranging from 15.08 - 46.46 dB, and the Structural SIMilarity index (SSIM) range from 0.89 to 1 for different patients' whole images and over lung areas. Our model is trained and tested using two novel COVID19 datasets that combined consist of 1049 COVID-19 positive cases.

### VI. FUTURE WORK

We show that it is possible for GANs to refine full resolution CT scan images based on conditional approximations, but ongoing future work is necessary before full-resolution synthetic 3D CT scan volumes can be generated from random noise. We propose an iterative upsampling approach of first generating a low resolution (pixelated) image with GAN and iterative depixelating the image with cGAN until we arrive at a full resolution 3D scan volume. We anticipate that this approach will enable the first full resolution CT scan volumes to be generated synthetically.

### ACKNOWLEDGMENT

This research was supported by NSF RAPID award titled: Deep Learning Models for Early Screening of COVID-19 using CT Images, award # 2027628. Authors would like to thank Dr. Eliot Siegel for his contributions.

### REFERENCES

[1] Y. Mirsky, T. Mahler, I. Shelef, and Y. Elovici, "CT-GAN: Malicious tampering of 3D medical imagery using deep learning," In 28th {USENIX} Security Symposium ({USENIX} Security 19) (pp. 461-478), 2019.

[2] N. E. M. Khalifa, M. H. N. Taha, A.E. Hassanien, and S. Elghamrawy, "Detection of coronavirus (COVID-19) associated pneumonia based on generative adversarial networks and a fine-tuned deep transfer learning model using chest X-ray dataset.", arXiv preprint arXiv:2004.01184, 2020.

[3] M. Jamshidi, A. Lalbakhsh, J. Talla, Z. Peroutka, F. Hadjilooei, P. Lalbakhsh, and A. Sabet, "Artificial Intelligence and COVID-19: Deep Learning Approaches for Diagnosis and Treatment", IEEE Access, 8, 109581-109595, 2020.

[4] P. Isola, J. Y. Zhu, T. Zhou, and A. A. Efros, "Image-to-image translation with conditional adversarial networks", In Proceedings of the IEEE conference on computer vision and pattern recognition (pp. 1125-1134), 2017.

[5] M. Mirza, and S. Osindero, "Conditional generative adversarial nets", arXiv preprint arXiv:1411.1784, 2014.

[6] S. Menon, J. Galita, D. Chapman, A. Gangopadhyay, J. Mangalagiri, P. Nguyen, and M. Morris, "Generating Realistic COVID19 X-rays with a Mean Teacher+Transfer Learning GAN", arXiv preprint arXiv:2009.12478, 2020.

[7] O. Gozes, M. Frid-Adar, H. Greenspan, P. D. Browning, H. Zhang, W. Ji, and E. Siegel, "Rapid ai development cycle for the coronavirus (covid-19) pandemic: Initial results for automated detection & patient monitoring using deep learning ct image analysis", arXiv preprint arXiv:2003.05037, 2020.

[8] A. Bernheim, X. Mei, M. Huang, Y. Yang, Z. A. Fayad, N. Zhang, and S. Li, "Chest CT findings in coronavirus disease-19 (COVID-19): relationship to duration of infection. Radiology", 200463, 2020.

[9] He, Xuehai, Xingyi Yang, Shanghang Zhang, Jinyu Zhao, Yichen Zhang, Eric Xing, and Pengtao Xie. "Sample-Efficient Deep Learning for COVID-19 Diagnosis Based on CT Scans",medRxiv, 2020.

[10] U. Sara, M. Akter and M. S. Uddin, " Image quality assessment through FSIM, SSIM, MSE and PSNR—a comparative study", Journal of Computer and Communications, 7(3), 8-18, 2019.

[11] Y. Song, S. Zheng, L. Li, X. Zhang, Z. Huang, and Y. Chong, " Deep learning enables accurate diagnosis of novel coronavirus (COVID-19) with CT images", medRxiv, 2020.

[12] F. Shan, Y. Gao, J. Wang, W. Shi, N. Shi, M. Han, and Y. Shi, "Lung infection quantification of covid-19 in ct images with deep learning", arXiv preprint arXiv:2003.04655, 2020.

[13] P. Nguyen, D. Chapman, S. Menon, M. Morris, and Y. Yesha, "Active semi-supervised expectation maximization learning for lung cancer detection from Computerized Tomography (CT) images with minimally label training data", In Medical Imaging 2020: Computer-Aided Diagnosis, Vol. 11314, p. 113142E, International Society for Optics and Photonics, 2020.

[14] "Italian Society of Medical and Interventional Radiology COVID19 dataset", SIRM, https://www.sirm.org/category/ senza-categoria/covid-19. Access date: August, 2020

[15] "COVID-19 CT segmentation dataset", https://medicalsegmentation.com/covid19/ . Access date: September, 2020

[16] "Radiopedia" https://radiopaedia.org/articles/ covid-19-4. Access date: September, 2020

[17] M. I. Vay et al., "BIMCV COVID-19+: a large annotated dataset of RX and CT images from COVID-19 patients," arxiv:2006.01174 2020

[18] S. P. Morozov, A. E. Andreychenko, N. A. Pavlov, et al., "MosMedData: Chest CT scans with COVID-19 related findings dataset," 2020. arXiv:2005.06465

[19]Xin Yi, Ekta Walia, and Paul Babyn. Generative adversarial network in medical imaging: A review. arXiv preprint arXiv:1809.07294, 2018.

[20] H. Uzunova, J. Ehrhardt, F. Jacob, A. Frydrychowicz,and H. Handels, "Multi-scale GANs for Memory-efficient Generation of High Resolution Medical Images", In International Conference on Medical Image Computing and Computer-Assisted Intervention (pp. 112-120). Springer, Cham, 2019.

[21]"Image quality assessment", https://www.sciencedirect.com/topics/engineering/image-quality-assessment#:~:text=The%20range%20of%20SSIM%20values,circularly%20symmetric%20Gaussian%20weighting%20function. Access date: November, 2020